\documentclass[lettersize,journal]{IEEEtran}

\usepackage{graphicx}
\usepackage{mathtools}
\usepackage{amsmath}
\usepackage{bm}

\usepackage{algorithm}
\usepackage{algpseudocode}

\usepackage{booktabs}
\usepackage{multirow}
\usepackage{makecell}
\usepackage{threeparttable}
\usepackage{siunitx}

\usepackage[table,dvipsnames]{xcolor}
\usepackage{pifont}
\usepackage{textcomp}
\usepackage{url}
\usepackage{lipsum}
\usepackage[acronym]{glossaries}
\makeglossaries

\usepackage{tikz}
\usepackage{pgf}
\usepackage{circuitikz}
\usetikzlibrary{arrows.meta, positioning, calc, decorations.markings, decorations.pathreplacing, shapes.geometric}
\usepackage{pgfplots}
\pgfplotsset{compat=1.18}

\usepackage{csquotes}
\usepackage{scalefnt}
\usepackage{wasysym}
\usepackage[hidelinks]{hyperref}
\usepackage{breakurl}
\usepackage[capitalise]{cleveref}

\usepackage{xspace}
\usepackage{todonotes}
\usepackage{soul}

\usepackage[backend=biber, style=ieee, citestyle=numeric-comp, hyperref, maxnames=10, minnames=1, giveninits=true, maxcitenames=2, mincitenames=1,bibencoding=utf8, doi=true, date=year, isbn=false]{biblatex}
\DeclareSourcemap{
  \maps[datatype=bibtex]{
    \map{
      \step[fieldset=issn, null]
    }
  }
}
\DefineBibliographyStrings{english}{
  andothers = {et al\adddot}
}

\newacronym{pe}{PE}{processing element}
\newacronym{ip}{IP}{intellectual property}
\newacronym{simd}{SIMD}{single instruction, multiple data}
\newacronym{ctc}{CTC}{computation-to-communication }
\newacronym{lstm}{LSTM}{long short-term memory}
\newacronym{lrcn}{LRCN}{long-term recurrent convolutional network}
\newacronym{cgra}{CGRA}{coarse-grained reconfigurable architecture}
\newacronym{gpu}{GPU}{graphics processing unit}
\newacronym{dse}{DSE}{design space exploration}
\newacronym{ml}{ML}{machine learning}
\newacronym{nn}{NN}{neural network}
\newacronym{dnn}{DNN}{deep neural network}
\newacronym{nre}{NRE}{non-recurring engineering}
\newacronym{tml}{TinyML}{tiny machine learning}
\newacronym{mcu}{MCU}{microcontroller unit}
\newacronym{hls}{HLS}{high level synthesis}
\newacronym{lut}{LUT}{look-up table}
\newacronym{nas}{NAS}{neural architecture search}
\newacronym{cnn}{CNN}{convolutional neural network}
\newacronym{fpga}{FPGA}{field-programmable gate array}
\newacronym{dsp}{DSP}{digital signal processing}
\newacronym{mac}{MAC}{multiply-accumulate}
\newacronym{asic}{ASIC}{application-specific integrated circuit}
\newacronym{rtl}{RTL}{register transfer level}
\newacronym{ai}{AI}{artificial intelligence}
\newacronym{cpu}{CPU}{central processing unit}
\newacronym{bram}{BRAM}{block random access memory}
\newacronym{ac}{AxC}{approximate computing}
\newacronym{msb}{MSB}{most significant bit}
\newacronym{os}{OS}{operating system}
\newacronym{ps}{PS}{processing system}
\newacronym{pl}{PL}{programmable logic}
\newacronym{dma}{DMA}{direct memory access}
\newacronym{icap}{ICAP}{internal configuration access port}
\newacronym{pcap}{PCAP}{processor configuration access port}
\newacronym{fp}{FP}{floating point}
\newacronym{dpu}{DPU}{deep-learning processor unit}
\newacronym{pr}{PR}{partial reconfiguration}
\newacronym{ff}{FF}{flip-flop}
\newacronym{qat}{QAT}{quantization aware training}
\newacronym{vww}{VWW}{visual wake words}
\newacronym{soc}{SoC}{system-on-chip}
\newacronym{dnndk}{DNNDK}{deep neural network development kit}
\newacronym{gemm}{GEMM}{general matrix multiplication}
\newacronym{gnn}{GNN}{graph neural network}
\newacronym{wmd}{WMD}{weight matrix decomposition}
\newacronym{iot}{IoT}{internet of things}
\newacronym{po2}{Po2}{power of two}
\newacronym{llm}{LLM}{large language model}
\newacronym{ga}{GA}{genetic algorithm}
\newacronym{mlp}{MLP}{multi-layer perceptron}
\newacronym{npu}{NPU}{neural processing unit}
\newacronym{bnn}{BNN}{binary neural network}
\newacronym{sa}{SA}{systolic array}
\newacronym{nsga}{NSGA-II}{non-dominated sorting genetic algorithm}
\newacronym{ptq}{PTQ}{post-training quantization}

\newcommand{\TODO}[1]{}            % disable TODO macro by default
\newcommand{\magenta}[1]{\textcolor{black}{#1}}

\colorlet{fhcolor}{black} %ProcessBlue
\newcommand{\fh}[2][]{}

\newcommand{\red}[1]{\textcolor{black}{#1}}
\newcommand{\blue}[1]{\textcolor{black}{#1}}
\newcommand{\orange}[1]{\textcolor{black}{#1}} %RedOrange
\newcommand{\green}[1]{\textcolor{black}{#1}} %ForestGreen

\newcommand{\cmark}{\color{OliveGreen}\ding{51}}
\newcommand{\xmark}{\color{BrickRed}\ding{55}}

\usepackage{amsmath,amsfonts}
\usepackage{array}
\usepackage[caption=false,font=normalsize,labelfont=sf,textfont=sf]{subfig}
\usepackage{textcomp}
\usepackage{stfloats}
\usepackage{url}
\usepackage{verbatim}
\usepackage{graphicx}
\def\BibTeX{{\rm B\kern-.05em{\sc i\kern-.025em b}\kern-.08em
    T\kern-.1667em\lower.7ex\hbox{E}\kern-.125emX}}
\begin{document}
\title{Co-Design of CNN Accelerators for TinyML using Approximate Matrix Decomposition%\\[-.25em]
}

\author{José Juan Hernández Morales*, Georgios Mentzos*, Frank Hannig, Konstantinos Balaskas, Georgios Zervakis, Jörg Henkel, and Jürgen Teich
\thanks{*These authors contributed equally to this work.}
\thanks{This work was partly supported by the Deutsche Forschungsgemeinschaft (DFG, German Research Foundation) under project number 524986327 (NA$^3$Os)}
\thanks{José Juan Hernández Morales, Frank Hannig, and Jürgen Teich are with the Department of Computer Science, Chair of Hardware/Software Co-Design, Friedrich-Alexander-Universität Erlangen-Nürnberg (FAU), Germany (e-mail: jose.juan.hernandez@fau.de; frank.hannig@fau.de; juergen.teich@fau.de).}%
\thanks{Georgios Mentzos and Jörg Henkel are with the Department of Computer Science, Chair for Embedded Systems, Karlsruhe Institute of Technology (KIT), Germany (e-mail: mentzos.george@kit.edu; joerg.henkel@kit.edu).}
\thanks{Konstantinos Balaskas is with the Computer Engineering Informatics Department, University of Patras, Greece (e-mail: kompalas@ceid.upatras.gr).}
\thanks{Georgios Zervakis is with the School of Electrical and Computer Engineering (ECE), National Technical University of Athens (NTUA), Greece (e-mail: zervakis@mail.ntua.gr)}
}

% \markboth{Journal of \LaTeX\ Class Files,~Vol.~18, No.~9, September~2020}%
% {How to Use the IEEEtran \LaTeX \ Templates}

\maketitle

\begin{abstract}

The paradigm shift towards local and on-device inference under stringent resource constraints is represented by the \gls{tml} domain. The primary goal of \gls{tml} is to integrate intelligence into tiny, low-cost devices under strict resource, energy, and latency constraints. However, the ultra-resource-constrained nature of these devices can lead to increased inference execution time, which can be detrimental in latency critical applications. 
At the same time, \gls{tml} applications are often associated with sensitive data. As such, latency optimization approaches that rely on training samples are infeasible when such data is unavailable, proprietary, or sensitive, highlighting a pressing need for optimization approaches that do not require access to the training dataset and can be applied directly to pre-trained models. Replacing costly multiplications with more hardware-efficient operations, such as shifts and additions, has been proposed as an effective method for reducing inference latency. However, post-training power-of-two (Po2) approaches are scarce and, in many cases, lead to unacceptable accuracy loss.

In this work, we propose a framework that applies approximate matrix decomposition to a given CNN in order to optimize hardware implementations subject to strict constraints and without any need of re-training or fine-tuning steps. The genetic algorithm-driven framework explores different matrix decompositions and resulting multiplier-less CNN accelerator designs for FPGA targets. A comprehensive evaluation of different \gls{tml} benchmarks demonstrates our framework's efficacy in generating latency-optimized implementations that satisfy strict accuracy and resource constraints, achieving an average \red{33\%} latency improvement with an average accuracy loss of \red{1.3\%} compared to typical systolic array-based FPGA accelerators.

\end{abstract}


\section{Introduction}

\IEEEPARstart{I}n recent years, the complexity of \gls{cnn} workloads has risen exponentially~\cite{steiner_olla_2022, xu_scaling_2018}, leading to a strong reliance on cloud-based processing.
Although cloud offloading provides scalability, it significantly strains computational resources, resulting in increased energy consumption and raising concerns about privacy and costs.
% \Gls{ai} applications are thus primarily offloaded to the cloud, raising concerns about rising carbon emissions, privacy and cost.
These challenges have driven a paradigm shift towards local, on-device, energy-efficient, and ultra-resource-constrained \gls{ml} deployment, which is represented by the \gls{tml} domain.
\Gls{tml} embeds intelligence in far-edge, cost-effective applications such as wearable healthcare monitoring~\cite{jia_tinyml_2024}, always-on anomaly detection~\cite{nassif_machine_2021}, and object detection~\cite{redmon_you_2016}.
% The advent of the \gls{tml} domain has brought intelligence to the far edge enabling applications such as wearable healthcare monitors~\cite{jia_tinyml_2024}, always on anomaly detection~\cite{nassif_machine_2021} and object detection~\cite{redmon_you_2016}.
Despite these advancements, \gls{tml} systems are highly resource-constrained in both memory and compute, with \glspl{mcu} or resource-constrained \glspl{fpga} widely employed as target platforms~\cite{noauthor_x-cube-ai_nodate,lai_cmsis-nn_2018,chen_tvm_2018,lin_mcunet_2020, mentzos2025r}. 
The inherent resource constraints of such devices can lead to increased inference latency, which can become critical in \gls{tml} applications such as tiny unmanned aerial vehicles~\cite{lamberti_distilling_2024} or wearable biomedical devices~\cite{turetta_lightweight_2025}, where inference must be performed under strict latency constraints. 
To enable deployment and reduce inference latency under such constraints, \glspl{cnn} are designed to be shallower~\cite{banbury_mlperf_2021} and architecturally simpler~\cite{howard_mobilenets_2017}, with a reduced number of parameters~\cite{he_deep_2016, iandola_squeezenet_2016}.
Techniques such as pruning and quantization have been widely adopted to reduce the hardware footprint of \gls{ml} models~\cite{han_deep_2016}.
Similarly, \gls{tml} acceleration has focused on developing hardware support for such techniques, including native computation at lower bitwidths~\cite{garofalo_xpulpnn_2021} or sparse processing~\cite{titopoulos_optimizing_2025}.
However, these approaches
% may incur considerable accuracy loss due to approximation.
% In addition, they 
still heavily rely on multiplication operations, which naturally introduce performance and resource overheads.
% However, these approaches may incur a considerable accuracy loss due to approximation, or may achieve suboptimal resource and power efficiency due to the use of costly multipliers.

% While techniques such as pruning and quantization~\cite{han_deep_2016} have been proposed as methods of further optimizing tiny \glspl{cnn} and have been adapted in accelerator both in industry and academia~\cite{noauthor_ethos-u55_nodate, chen_eyeriss_2019, jouppi_-datacenter_2017, ng_high_2024} but these still rely on \glspl{pe} comprised of costly multipliers.  

% \Glspl{fpga} are a promising platform for multiplier-less inference due to their arbitrary precision and dataflows.
% Additionally, \glspl{fpga} can -in contrast to \glspl{mcu} or \glspl{gpu}- be adapted to entirely avoid costly multiplications, using either binary or shift and add operations. 
% Particularly promising are \gls{po2} architectures, which have shown potential for significant improvements in terms of hardware efficiency~\cite{you_shiftaddnet_2020}. For example, in our experiments, a shift operation was shown to use $1.5x$ less \glspl{lut} on an \gls{fpga}, compared to a multiplier at an equivalent precision of 8-bits.
Multiplier-less inference has emerged as a promising strategy, replacing resource-demanding multiplications with lightweight operations, such as shift-and-add.
\glspl{fpga} are well-suited for this approach, offering design customizability and optimal support for such architectures.
Our initial experimentation, on a rather small \gls{fpga}, shows that a variable \green{signed right-shift} operation\fh{Do you mean a barrel shifter?} requires $1.5\times$ fewer \glspl{lut} than an equivalent multiplier, showcasing the potential resource savings of multiplier-less inference.
\Glspl{bnn}~\cite{zhang_fracbnn_2021, wang_lutnet_2020, umuroglu_finn_2017} and \gls{po2}-based networks~\cite{you_shiftaddnet_2020, lehnert2023most, muller2023linear, muller2021, elhoushi_deepshift_2021, zhou_incremental_2017} are notable examples, with the latter offering greater representation flexibility and potentially leading to higher achievable accuracy~\cite{you_shiftaddnet_2020}.
% However, \gls{po2} transformation often incurs substantial accuracy degradation, necessitating retraining to recover.
However, \gls{po2} transformation often necessitates training in order to avoid any substantial accuracy degradation.
As user privacy is paramount in \gls{tml} \textcolor{fhcolor}{scenarios}, retraining may be infeasible due to sensitive, proprietary, or unavailable training data, \textcolor{fhcolor}{which is typically the case in third-party or licensed models~\cite{aanjankumar2025enhanced}.
Rather, post-training approaches enable a rapid deployment and iteration without incurring the computational and energy costs of retraining, thereby reducing development overhead. 
% Finally, by avoiding retraining on limited or proxy datasets, post-training approaches mitigate the risk of overfitting and preserve the original model's generalization capabilities.
}
% However, the adaptation of multiplier-less based \gls{cnn} inference incurs significant accuracy loss, which typically requires training to recover. This can be impractical in applications where the training dataset is unavailable, proprietary of sensitive.
On the other hand, current \gls{po2} approaches that operate entirely in post-training scenarios~\cite{lehnert2023most, gudovskiy_shiftcnn_2017} are hardware-unaware, adhering to only accuracy constraints and neglecting the inherent resource constraints of \gls{tml}.
\emph{To the best of our knowledge, no existing work enables \gls{po2}-based \gls{cnn} inference under strict resource and accuracy constraints without retraining.} %\TODO{Emphasize retraining restrictions as motivation for post-training approach }
% These approaches, usually instantiate an accelerator considering only an accuracy constraint. This can lead to cases, where the accuracy constraint is satisfied but the final accelerator violates resource constraints, especially so in the \gls{tml} domain. \textit{So far there is no work that can enable \gls{po2} \gls{cnn} inference under tight resource constraints, without the need for training}. 

In this work, we address these limitations by proposing the first post-training, hardware-aware co-design framework for \gls{po2}-based \gls{cnn} inference on \gls{tml} devices.
Our framework leverages approximate \emph{\glsentrylong{wmd}} (\glsentryshort{wmd}) to transform targeted \glspl{cnn} into \gls{po2} form without the need for retraining, jointly targeting minimal latency and accuracy loss.
To that end, we design a programmable, multiplier-less \gls{cnn} accelerator tailored to the decomposed \gls{cnn}.
The accelerator replaces traditional \gls{mac} units with lightweight shift-and-add \glspl{pe} and is fully parametrizable based on our framework’s decomposition outputs.
Our framework jointly explores \gls{po2} transformation strategies and accelerator configurations. It uses our hardware-driven latency and resource analytical models to generate accuracy-latency Pareto-optimal solutions, subject to the resource constraints of \gls{tml}.
Evaluated on the established MLPerfTiny benchmark suite~\cite{banbury_mlperf_2021} and the \texttt{Arty A7-100T} \gls{fpga} board, our framework demonstrates an average latency improvement of \red{33\%} at just \red{1.3\%} average accuracy loss, compared to modern \magenta{8-bit accelerator} designs.\\[.5em]
\noindent
\textbf{Our novel contributions in this work are as follows:}
\begin{itemize}
    \item 
    We introduce the first post-training, hardware-aware co-design framework that jointly optimizes \glspl{cnn} and \gls{po2}-based accelerator configurations for Pareto-optimal trade-offs between accuracy and latency.
    \item 
    We design a parametrizable, multiplier-less \gls{fpga} accelerator with shift-and-add \glspl{pe} for efficient hardware deployment of \glspl{cnn}.
    \item 
    Our approach demonstrates substantial latency gains at negligible accuracy loss compared to the state of the art, while adhering to the strict constraints of \gls{tml}.
\end{itemize}

% \section{Motivation and Basic Terminology (FAU \& KIT) 1 page} 

\section{Background and Related Work}\label{sec:background}

\textcolor{fhcolor}{In this section, we first provide the fundamentals for decomposing a given CNN weight matrix into an approximated power-of-two form in \Cref{subsec:wmd_algorithm}, followed by a discussion of related work in \Cref{subsec:related}.}

\subsection{Approximate Weight Matrix Decomposition}\label{subsec:wmd_algorithm}

% Our proposed \frameworkdecomposition{} framework replaces multiplications with shift-add operations to reduce latency and meet \gls{tml}'s resource constraints.
% Thus, we adopt approximate \gls{wmd} for transforming \gls{cnn} weights into \gls{po2} form, using the approach proposed in~\cite{muller2023linear}.

\begin{figure}[t]
\centerline{\includegraphics[width=\columnwidth]{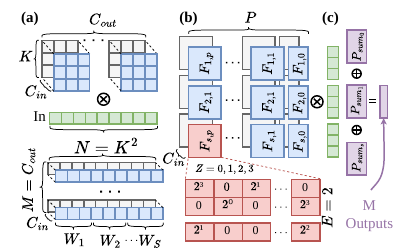}}
\caption{(a) CNN weight matrix transformation to $N$ and $M$ dimensions.
(b) WMD of the transformed weight matrix to P decomposed F matrices with Po2 weights.
(c) WMD multiplication with input and accumulation of the partial sums.
\TODO{FH: In the figure, use the same \emph{math font} as throughout the paper, that means, e.g., $C_{out}$ instead of $\text{C}_{\text{out}}$, $N$ instead of $\text{N}$, $W_{s}$ (intex is lower case $s$) instead of $\text{W}_{\text{S}}$ (here uppercase), $F_{s,P}$ instead of $\text{F}_{\text{S,P}}$, and so on.}}
\label{fig:wmd}
\end{figure}

In~\cite{muller2023linear} a mathematical weight decomposition technique is presented, which transforms \gls{cnn} weights into \gls{po2} form.
The goal of the \gls{wmd} process is, given a weight matrix $W$, to obtain an approximate representation $\hat{W}$ where the multiplications have been replaced by shift-and-add operations.
The first step of the process can be seen in~\cref{fig:wmd}a. 
The weight matrix of a \gls{cnn} layer is first flattened in the kernel dimension $K^2$ to form the $N=K^2$ dimension, with the output channels $C_{out}$ concatenated to form $M$.
After the $M \times N$-matrix $W$ is constructed, it is then split into a set of slices $\mathcal{S}$ with a slice width of $S_W$, such that $W = [W_{1} | W_{2} | ... |W_{s}]$.

The decomposition process of each $W_s^{M \times S_W}$ matrix is illustrated in~\cref{fig:wmd}b.
Each $W_s$ can be approximated using a set of $\mathcal{F}^{M \times M}$ matrices containing a total of $P$ decomposed matrices $F$, such that $W_s \approx F_{s,P} \ldots F_{s,1} F_{s,0}$. 
The initial matrix $F_{s,0}^{M \times S_W}$ has the same dimensions as $W_s$ and is composed of an identity matrix $I^{S_W \times S_W}$ and then padded with $(M-S_W)$ zeroes.
Then, each decomposed array $F$ is constrained to contain exactly $E$ non-zero elements, each with a range of $\pm 2^Z$ \gls{po2} values, where $Z\in \mathbb{Z}$ represents the number of possible shifts. 
This \gls{wmd} process can be fully described by the following parameters: $\{{P,Z,E,M,S_W}\}$, each one distinctly affecting the accuracy of the decomposed \gls{cnn}.

Finally, \cref{fig:wmd}c presents the decomposed version of a \gls{cnn} convolution operation, assuming a convolutional kernel of size $N$ sliding over the input feature map. 
An input vector $In$ of size $N$ is also split into $S$ slices, such that $In = [In_{1} | In_{2} | ... |In_{s}]$. A slice $In_s$ is then applied upon the decomposed matrices $F$ (corresponding to $W_s$) to produce $S$ partial sums.
The convolution operation is then completed by accumulating the partial sums, as generated by the decomposed convolution, to produce $M$ outputs. 

% The above formulation provides a rigorous methodology for transforming \gls{cnn} weights into \gls{po2} values, enabling resource-efficient inference within \frameworkdecomposition{}. 

\subsection{Related Work}\label{subsec:related}
\gls{cnn} inference acceleration has been widely studied, with numerous works proposing specialized compute engines, mostly based on 8-bit \gls{mac} units~\cite{chen_eyeriss_2019, jouppi_-datacenter_2017}.
Specifically for \gls{tml}, ARM's Ethos-U55~\cite{noauthor_ethos-u55_nodate} is an embedded \acrshort{npu} targeting \glspl{mcu}, supporting block sparsity and hardware support for a minimum precision of 8 bits. 
Similarly, \citeauthor{titopoulos_optimizing_2025}~\cite{titopoulos_optimizing_2025} integrated structured sparsity into vector execution on resource-constrained RISC-V cores to accelerate matrix multiplications.
\citeauthor{ng_high_2024}~\cite{ng_high_2024} presented a streaming accelerator with layer-specific, mixed-precision \glspl{pe}.
\citeauthor{garofalo_xpulpnn_2021}~\cite{garofalo_xpulpnn_2021} provided native support for low-bitwidth execution of \gls{tml} workloads on \glspl{fpga}.
While these approaches optimize performance, they remain reliant on costly multipliers, which dominate resource utilization on constrained platforms.
% In terms of \gls{cnn} inference acceleration there has been significant research interest over the years. The works of \cite{chen_eyeriss_2019, jouppi_-datacenter_2017} proposed compute engines based on 8-bit \gls{mac} operations. Ethos-U55~\cite{noauthor_ethos-u55_nodate} is an NPU developed by ARM targeting embedded inference on \glspl{mcu}. It has support for block sparsity as well as a minimum of 8-bits of precision. The authors of~\cite{ng_high_2024} propose a streaming architecture with layer-specific, mixed precision \glspl{pe}. However, these acceleration methodologies are primarily based on costly multipliers and therefore incur significant resource overhead. 

Multiplier-less inference has been explored through \glspl{bnn}~\cite{zhang_fracbnn_2021, wang_lutnet_2020, umuroglu_finn_2017} and \gls{po2} quantization~\cite{you_shiftaddnet_2020, lehnert2023most, muller2023linear, muller2021, elhoushi_deepshift_2021, zhou_incremental_2017}, with \gls{po2} methods offering more flexibility and potentially higher accuracy~\cite{you_shiftaddnet_2020}.
In~\cite{zhou_incremental_2017}, \citeauthor{zhou_incremental_2017} applied quantization-aware training to transform CNN weights into \gls{po2} values, while in DeepShift~\cite{elhoushi_deepshift_2021}, \citeauthor{elhoushi_deepshift_2021} replaced multiplications with bitwise shifts and sign flips for both inference and training.
\citeauthor{you_shiftaddnet_2020} introduced ShiftAddNet~\cite{you_shiftaddnet_2020}, a model that separates shift and add layers to enhance accuracy.
\emph{However, these techniques rely on retraining, which is infeasible when training data are proprietary, sensitive, or unavailable.}  
% In terms of \gls{po2} based inference, the authors of \cite{zhou_incremental_2017} quantized \gls{cnn} weights to \gls{po2} followed by re-training to recover accuracy loss. In~\cite{elhoushi_deepshift_2021} multiplications were replaced by bit-wise shits and sign flips during both inference and training. ShiftAddNet~\cite{you_shiftaddnet_2020} introduces separate layers for shift and add operations for both training and inference to improve accuracy. \textit{However, these methodologies are impossible to implement in cases where the training dataset is unavailable}

Several \gls{po2}-based approaches avoid retraining entirely.
\citeauthor{muller2021}~\cite{muller2021} introduced \gls{wmd} to approximate matrices using sparse and \gls{po2} components.
\citeauthor{lehnert2023most}~\cite{lehnert2023most} proposed a post-training \gls{po2} quantization technique, used to implement an \gls{fpga} accelerator for small models with fully-unfolded data\-paths and hardwired shifts and zeros.
This was later extended to \glspl{cnn} by \citeauthor{muller2023linear}~\cite{muller2023linear}.
With ShiftCNN~\cite{gudovskiy_shiftcnn_2017}, \citeauthor{gudovskiy_shiftcnn_2017} proposed a \gls{po2} quantization algorithm alongside a \gls{po2}-based \gls{fpga} accelerator, while \citeauthor{you_shiftaddllm_2024}~\cite{you_shiftaddllm_2024} presented a multi-objective post-training reparametrization to minimize quantization error.
\emph{However, these methods do not explicitly account for hardware resource constraints, often leading to designs that exceed the strict budget of TinyML devices and necessitate using large or even server-grade \gls{fpga} boards, making them impractical for far-edge deployment.}
% On the other hand, the authors of~\cite{muller2021} used \gls{wmd} to approximate a given matrix by a series of sparse and \gls{po2} matrices. The methodology in~\cite{lehnert2023most} was used to build an \gls{fpga} accelerator for simple \glspl{mlp}. The proposed architecture was fully unfolded with hardwired shift and zero elements. In~\cite{muller2023linear} the approach was further extended to support \glspl{cnn}. ShiftCNN~\cite{gudovskiy_shiftcnn_2017} proposed a \gls{po2} quantization methodology alongside a \gls{po2} \gls{fpga} accelerator. Finally, the authors of \cite{you_shiftaddllm_2024} proposed a post-training multi-objective reparametrization to minimize weights and activation error. \textit{However, these do not jointly consider in their implementation, hardware limitations making them in-applicable to the \gls{tml} domain.}

Our proposed framework addresses these gaps by jointly optimizing \gls{cnn} weight decomposition and accelerator design to enable efficient, resource-compliant \gls{po2} inference for \gls{tml} devices without retraining.  
In \cref{tab:related}, we present a qualitative evaluation of our techniques, comparing them with state-of-the-art \gls{cnn} inference approaches.
% Our proposed \frameworkdecomposition{} framework aims to address these limitations and provide a co-design methodology of accelerating tiny \glspl{cnn} using \gls{po2} representation.
% \frameworkdecomposition{} can collaboratively explore \gls{cnn}-accelerator pairs to provide accurate \gls{po2} inference, while adhering to hardware and accuracy constraints without the need for training.  

\begin{table}
\setlength\tabcolsep{4pt}
\centering
% \scriptsize
\caption{Qualitative taxonomy of the differentiating factors between our proposed approach and the state of the art.}
\begin{threeparttable}
\begin{tabular}{c|ccccc}

    \toprule
    \textbf{Approach} & 
    {\thead{\textbf{Multiplier-less}\\\textbf{Inference}}} &
    % \textbf{Post-Training} &
    {\thead{\textbf{Without}\\\textbf{Retraining}}} &
    % \textbf{Co-Exploration}
    {\thead{\textbf{Hardware}\\\textbf{Aware}}}
    \\ 
    \midrule

    % quantized accelerators
    \cite{noauthor_ethos-u55_nodate, chen_eyeriss_2019, jouppi_-datacenter_2017, ng_high_2024} &
    \xmark & 
    \cmark & \cmark
    % & \hcircle
    \\
    % \hline

    % training & po2
    \cite{you_shiftaddnet_2020, elhoushi_deepshift_2021, zhou_incremental_2017} & 
    \cmark &
    \xmark & \cmark
    % & \ecircle  
    \\ 
    % \hline

    % post training & po2
    \cite{muller2021, lehnert2023most, muller2023linear, gudovskiy_shiftcnn_2017} &
    \cmark &
    \cmark & \xmark
    % & \fcircle 
    \\ 
    % \hline

    \midrule
    \textbf{Ours} &
    \cmark &
    \cmark & \cmark
    % & \fcircle
    \\ 
    \bottomrule

\end{tabular}
\end{threeparttable}
\label{tab:related}
\end{table}

\section{Proposed Programmable Accelerator}
\label{sec:accelerator}

We propose a programmable \gls{cnn} accelerator for \gls{tml} deployment on resource-constrained \glspl{fpga}.
Our accelerator comprises a \gls{sa}, i.e., a 2-D grid of lightweight \glspl{pe}, capable of supporting the \gls{wmd} algorithm detailed in~\cref{subsec:wmd_algorithm} and performing exclusively shift-and-add operations. 
Its programmability allows flexible mapping of computations, including folding workloads across multiple passes when necessary due to resource constraints. 
The details of the \gls{pe} are described in Section~\ref{subsec:pe}, and the \gls{sa} arrangement and scheduling are presented in Section~\ref{subsec:systolic}.

\subsection{Processing Element \textcolor{fhcolor}{(PE)}}
\label{subsec:pe}

The microarchitecture of our \gls{pe} is shown in~\cref{fig:accelerator_scheduling}a. Each \gls{pe} contains \magenta{pipelined} sub-modules (referred to as F-blocks hereafter), 
% \todo[inline]{Now not clear how P is defined. F-blocks not defined explicitly in figure} 
computing the multiplication of an input vector $V_{in}$ by an ${F}$ matrix.
In our \gls{pe}, we implement $F_0$ and $F_{gen}$ as hard blocks, meaning that in the case of $P>2$, the $F_{gen}$ block is time-multiplexed. The \gls{pe} computes the multiplication of one slice of the decomposed matrix as described in~\cref{subsec:wmd_algorithm}, computing the equivalent of $S_W$ convolution operations per row, for a total of $M$ rows. Considering that each matrix ${F}$ is restricted to have exactly $E$ non-zero elements per row, each row of the F-block comprises $E$ shift units, which are then reduced using an adder tree. Moreover, the sparsity within the ${F}$ matrices is unstructured, meaning that each shift unit also requires a multiplexer to select which element of the $V_{in}$ is passed to the shift unit.
The shift unit itself  is comprised of $Z$ predefined shift values, which are selected using multiplexers.

Moreover, we take advantage of some observed properties of the ${F}$ matrices:
First, ${F}_{s,1}$ matrices always contain their non-zero elements in their first $S_W$ columns, which allows for omitting position encoding bits in the signals and multiplexers before the multiplier inputs.
Second, in an effort to reduce indexing cost, we replace one of the non-zero elements $E$ with a fixed value of $1$ on the diagonal of the $F$ matrix (referred to as diagonal optimization hereafter). The activation is then passed directly to the adder tree, which allows for encoding $E-1$ elements per row. Then, all F-blocks corresponding to these matrices just need $E-1$ multiplier units per row, while each element of the input vector is hardwired to the adder tree of the corresponding row. Finally, for the purpose of lowering hardware resources, our design considers a set of \gls{po2} with only negative exponents, which allows implementing right-shifts only. 
% Finally, our design considers a set of powers of 2 with just negative exponents, defined as $\{\pm2^{-k} | k \in \mathbb{Z},  0 \leq k \leq Z\}$ for the purpose of lowering hardware resources, which corresponds to implementing right-shifts only.
% \todo[inline]{clarify it is just rightshift}

\subsection{Systolic Array \textcolor{fhcolor}{(SA)}}
\label{subsec:systolic}

\begin{figure}
\centerline{\includegraphics[width=\columnwidth]{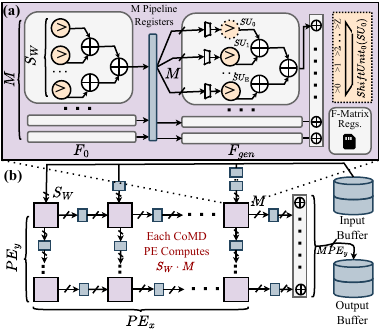}}
\caption{(a) Microarchitecture of our \gls{wmd} \gls{pe}. (b) Our \glsentrylong{sa} (\glsentryshort{sa}), composed of pipelined \glspl{pe}, handling $PE_x \cdot PE_y \cdot S_W \cdot M$ convolution operations every cycle. \TODO{Split and have a) at top and b) at bottom}}
% \caption{(a) Microarchitecture of our \gls{pe}, which consists of $F_0$ and $F_{gen}$. Po2-based multiplications are handled by the shift unit, and the output is accumulated with the results from the previous \gls{pe}. F-Matrices are stored in registers within the \gls{pe}. Each \gls{pe} computes $S_W \cdot M$ convolution operations. (b) Our \gls{sa} composed of pipelined \glspl{pe}. The inputs are fed to the \gls{sa} via the input buffer and the outputs from each row are accumulated and then stored in the output buffer. Our \gls{sa} handles $PE_x \cdot PE_y \cdot S_W \cdot M$ convolution operations every cycle.}
\label{fig:accelerator_scheduling}
\end{figure}

Our \glspl{pe} are arranged in a 2-D grid in a pipelined manner, forming the core \gls{sa} of our accelerator architecture, as shown in \cref{fig:accelerator_scheduling}b.
% Using the \gls{pe} architecture defined in~\cref{subsec:pe} we define our accelerator as a \gls{sa}, with the \glspl{pe} pipelined in a 2-dimensional grid as can be seen in~\cref{fig:accelerator_scheduling}b.
To map a convolution operation to our accelerator, we follow a process similar to~\cref{subsec:wmd_algorithm}, but substituting $N$ for $C_{in}$.
As can be seen in~\cref{fig:accelerator_scheduling}b, the input features reside in an on-chip input buffer, and then $S_W$ inputs flow from the top of the \gls{sa}.
Since each row corresponds to different output channels, the input features are shared along the y-axis.
The partial sums are reduced along the x-axis via internal accumulators within each \gls{pe}.
They are then summed up with the outputs from the previous iteration and stored in an on-chip output buffer.
The \glspl{pe} in the x-axis are connected using $M$ adders.\fh{Is it $M$ adders? Because for adding $M$ values, you require $M-1$ addition? Answer: These M values are not accumulated. These adders just M parallel additions from current PE and previous PE results.}
The parallelism of our \gls{sa} (i.e., useful convolution\fh{fuzzy, be more precise what a \enquote{useful conv.\ MAC} is} \glspl{mac} per cycle) can be determined based on the total number of \glspl{pe} and the parallelism within each individual \gls{pe}. At the \gls{pe} level, this depends on the \gls{wmd} parameters $S_W$ and $M$.
Then the total number of \green{outputs equivalent to} \glspl{mac}\fh{What is \enquote{useful}? Answer: just removed the word useful} per cycle of our accelerator is $S_W \cdot M \cdot PE_x \cdot PE_y$, with $PE_x$ the number of columns and $PE_y$ the number of rows of the \gls{sa}.

\begin{figure*}[b!]
\centerline{\includegraphics[width=\textwidth]{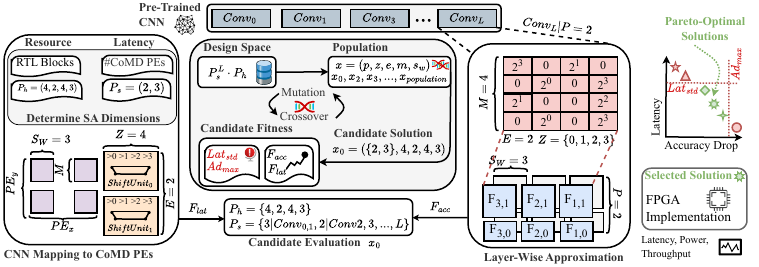}}
\caption{Our framework methodology. The input is a pre-trained \gls{cnn}, the design space, and additional accuracy and latency constraints. Candidate solutions are evaluated in terms of accuracy and latency, resulting in a \green{front of Pareto-optimal solutions}.}
\label{fig:frameworkheterogeneous}
\end{figure*}

\begin{algorithm}[t]
\caption{\gls{pe} Mapping}
\label{alg:pe_mapping}
\begin{algorithmic}[1]
\Require $R_{PE_{unit}}$, $LUT_{max}$
\State $L_{best} \gets \infty$
\State $M_{best} \gets None$
\State $PE_x\gets 1$  
\While{$R_{PE_x} \leq LUT_{max}$}
    \State $PE_y \gets \left\lfloor {LUT_{max}} / {R_{PE_x}} \right\rfloor$
    \State $Map_{PE} \gets (PE_x,PE_y)$
    \State $Lat \gets Lat_{accl}(Map_{PE})$
    \If{$Lat < Lat_{best}$}
        \State $Map_{best} \gets Map_{PE}$
        \State $Lat_{best} \gets Lat$
    \EndIf
    \State $PE_x\gets PE_x+1$ 
\EndWhile
\State \Return{$Map_{best}$, $Lat_{best}$}
\end{algorithmic}
\end{algorithm}

To minimize data movement, we fetch the elements of all ${F}$ matrices from \blue{on-chip} memory into registers within the \glspl{pe}, and then traverse through the \blue{input features} to be multiplied. This approach is analogous to the weight-stationary scheduling used in other \gls{sa}-based \gls{cnn} accelerators (e.g., TPU~\cite{jouppi_-datacenter_2017}).
For the loading of the corresponding ${F}$ matrices, a total of \magenta{$PE_x \cdot PE_y \cdot P \cdot E \cdot M$} elements need to be fetched and stored in registers.
Our \gls{sa} makes use of \gls{bram} resources available in \glspl{fpga} for the input and output buffers. One \gls{bram} is used to feed each column and ${PE_y \cdot M \cdot out_{bw}}/{b_{ports}}$ \glspl{bram} to store the packed results from all rows of the grid, with $out_{\textit{bw}}$ the bit-width of each output element and $b_{\textit{ports}}$ the summed bit-width of all ports of the \gls{bram}.

To determine the 2D mapping $(PE_x,PE_y)$ on our \gls{sa}, we use~\cref{alg:pe_mapping}.
Starting from $PE_x = 1$, the algorithm iteratively increments the length of the row, calculates $PE_y$ based on the total number of \glspl{lut} of the \gls{fpga}, and computes the latency $Lat$, based on the hardware cost of a single \gls{pe} unit $R_{PE_{unit}}$ and the cost of a \gls{pe} row $R_{PE_x}$, which can be derived from the F-blocks cost $R_{F_{gen}}$ and $R_{F_{0}}$ as described in~\cref{subsec:accelerator_modelling}. The process continues until the cost of one row exceeds the available \glspl{lut}.
Once the algorithm terminates, the best latency $Lat_{best}$ is returned, ensuring that for a given \gls{cnn}, the assignment best suited to its characteristics is chosen. This ensures that the best \gls{pe} mapping is chosen for a given \gls{cnn} based on the $C_{out}, C_{in}$ dimensions.

\section{Proposed Framework}
\label{sec:co-exploration}

In this section, we describe our novel hardware-aware framework for co-designing \gls{po2}-based \glspl{cnn} and dedicated accelerators.
Figure~\ref{fig:frameworkheterogeneous} presents an overview of our framework.
It takes as input a pre-trained \gls{cnn}, the design space of \gls{wmd} parameters for each layer, and predefined application constraints for accuracy and latency.
Then, candidate decomposed \glspl{cnn} and accelerators are explored---aided by our fully parametrizable architecture (see \cref{sec:accelerator})---employing high-level surrogate models for latency and resource utilization for the targeted \gls{fpga}.
Our framework outputs a Pareto front of \gls{po2}-decomposed \gls{cnn}-accelerator configurations with optimal latency-accuracy trade-offs, without requiring any additional training. 

% latency is directly affected by hw, is this clear from the intro ?
% \begin{figure}
% \centerline{\includegraphics[width=\columnwidth]{figures/weight_matrix_decomposition-codesign_methodology.pdf}}
% \caption{Our \frameworkdecomposition{} methodology. The input is a pre-trained \gls{cnn}, the design space, and the accuracy and latency constraints. Candidate solutions are evaluated in terms of accuracy and latency, resulting in an optimal Pareto front.}
% \label{fig:overview}
% \end{figure}

\subsection{Design Space} 
\label{subsec:dse}

% \todo[inline,color=blue!15]{\small $S_{w}$ vs. $S_{W}$, unify!\\[.5em]
% $Z$ is not introduced in \cref{subsec:wmd_algorithm}, only in a later section\\[.5em]
% \enquote{a polynomial dependency on $L$} What is $L$? (latency or no.\ of layers ?)\\
% Ambiguity:\\
% -- \enquote{best latency $L_{best}$}\\
% -- \enquote{each layer $L$}
% }

The \gls{wmd} algorithm is fully parametrized, given the following variables: $P, Z, E, M, S_{W}$ (see~\cref{subsec:wmd_algorithm}). Thus, a design space of different \gls{wmd}-related choices can be defined, each affecting the accuracy, latency, and resources of the \gls{cnn} accelerator.
% We can therefore define a co-design problem, allowing us to jointly optimize both the accuracy and latency of \gls{cnn} inference, using \frameworkdecomposition{}. 
The set of \gls{wmd} parameters is partitioned into a set of hard accelerator parameters $P_h$---applied to all decomposed layers $L$---and a set of soft, layer-specific choices $P_s$ applied individually to each different layer, setting a design space of cardinality $|P_s^{L}| \times |P_h|$. We choose to partition the design space in this way, primarily because of its complexity. For a given \gls{cnn}, the number of layers to be decomposed is constant; therefore, \magenta{it has a polynomial dependency on $L$}. Using this split, we can therefore have control over the complexity of the design space. Secondly, allowing each layer to be decomposed with $P_s$ \gls{wmd} parameters means \magenta{that the hardware needs to be able to support all choices made by each layer $L$, which can cause significant overhead in our programmable accelerator}. In short, partitioning the design space in this manner allows us to control the cost in terms of hardware as well as the size of the design space. Although different partitions of \gls{wmd} parameters can be considered, for the rest of this paper we set the number of decomposition blocks to be the only soft parameter ($P_s = \{P\}$) and the rest to be hard parameters $P_h = \{Z, E, M, S_{W}\}$.\fh[inline]{By this separation into soft and hardware parameters, which restricts the search space, can you guarantee that there are no \emph{better} solutions that are also Pareto-optimal or even dominate points on your Pareto front?}

% \todo[inline,color=blue!15]{\small Are $P_s$ and $P_h$ sets? Then, I would write: $P_s = \{P\}$ and $P_h = \{Z, E, M, S_{W}\}$.}

% mention that the pe mapping is part of the exploration ????? 

\subsection{Accelerator Modeling} 
\label{subsec:accelerator_modelling}

\subsubsection{Resource Model}
\label{subsec:resource_model}
The total resources \blue{$R_{accl}$ for our accelerator, in terms of FPGA LUTs,} can be estimated using~\cref{eq:cost_dec}. \blue{This value is} dependent on the resources of $F_0$ and $F_{gen}$, the total number of \glspl{pe} $PE_x$, $PE_y$, as well as the reduction cost between the \glspl{pe} in the x-dimension. 

\begin{equation}
    R_{accl} = PE_y PE_x (R_{F_{0}} + {R_{F_{gen}}}  + R_{add} M)
    \label{eq:cost_dec}
\end{equation}

The resources for an F-block are detailed in~\cref{eq:cost_fgen} and~\cref{eq:cost_f1}. We begin our resource analysis from the resources $R$ 
% \todo[inline]{introduce $R$ before equation}
of the base units used within our \glspl{pe}.
Then, we estimate the resources of our \gls{pe} by multiplying the base cost by the number of times each block is instantiated. As described in~\cref{subsec:pe} for one F-block, we use a \gls{po2} shift unit, a multiplexer, and an adder tree with base resources of $R_{mul}$, $R_{mux}$, which are dependent on the supported shifts $Z$ and the bit-width of \blue{the maximum supported \gls{wmd} stage $F_{max}$, respectively}. Similarly, $R_{add}$ is dependent on the number of $E$ elements to be reduced and $F_{max}$.
Moreover, we also account for the simplifications of the first F-block $F_0$ and the diagonal optimization in $F_{gen}$.
Note that our resource models use direct hardware measurements, obtained from EDA tools for the target \gls{fpga}, and thus provide highly accurate estimations guiding the optimization process.

\begin{equation} \label{eq:cost_fgen}
\begin{split}
R_{F_{gen}}(E,F_{max},Z) = 
M(E-1) (R_{mul} + R_{mux}) + R_{add})
\end{split}
\end{equation}
\begin{equation}
    \begin{split}
    R_{F_0}(S_W,F_{0},Z) = 
    M(S_{W} R_{mul} + R_{add})
    \end{split}
    \label{eq:cost_f1}
\end{equation}

\subsubsection{Latency Model}
\label{subsec:latency}

As described in \cref{subsec:systolic}, our accelerator offers parallelism in two distinct ways with a total parallelism of $(S_W \cdot PE_x, M \cdot PE_y)$ in the x- and y-dimension, respectively.
We estimate the latency of the accelerator based on the total operation count within each convolution.
This is dependent on the size of the output $O_{x,y}$, the kernel size $K_{x,y}$, and the input and output channels $C_{in}$ and $ C_{out}$.
$C_{in}$ is parallelized along the x-dimension and $C_{out}$ along the y-dimension.
The final latency of the accelerator can be derived by dividing $C_{in}$ and $C_{out}$ by the total parallelism in the x- and y-dimension, respectively.
% \blue{Considering the additional latency introduced when changing the $F$ matrices values within the PEs registers, which includes the writing and the start of continuous outputs at all the rows of the systolic array, a constant $Lat_{w}=2 \cdot PE_x + 2 \cdot PE_y$ is added each full traversal of the input feature map. A factor of 2 is used for $PE_x$ since a pipeline stage is used for each F-block, and the same factor is used for $PE_y$ since one cycle per row is needed to fill all registers (design decision to avoid BRAM bandwidth limit).}
Additionally, as detailed in~\cref{subsec:systolic}, 
% \todo[inline]{not detailed} 
our accelerator has support for the $F_0$ and a generic block $F_{gen}$, meaning that for $P$ total decompositions, there is hardware support for implementing the first two blocks $F_0,F_{gen}$ in \magenta{one cycle}, while the rest is time-multiplexed over $F_{gen}$.
This results in an additional latency factor of $Lat_F=1+(F_{max}-\magenta{2})$. The total accelerator latency can therefore be derived by summing the individual layer latencies.

% \begin{equation}
%     Lat_{accl} = \sum_{l=1}^{L}{Lat_{F} ({K_{x,y}}  {O_{x,y}} + Lat_w) \left\lceil \frac{C_{in}}{S_w \cdot PE_x}\right\rceil} \left\lceil\frac{C_{out}}{M \cdot PE_y}\right\rceil
%     \label{eq:latency_model}
% \end{equation}

\begin{equation}
    Lat_{accl} = \sum_{l=1}^{L}{Lat_{F} {K_{x,y}}  {O_{x,y}} \left\lceil \frac{C_{in}}{S_W \cdot PE_x}\right\rceil} \left\lceil\frac{C_{out}}{M \cdot PE_y}\right\rceil
    \label{eq:latency_model}
\end{equation}
\TODO{GM : Fix eq 4}

\subsection{Genetic-Based DSE}
% consider clarifying with steps sampled solution --> resources --> how many PEs --> latency, contraints (accuracy , latency better than std latency )
The core of our framework is a hardware-aware two-objective optimization process, aiming to balance \gls{cnn} accuracy with inference latency in a search space that is additionally restricted by the strict constraints of a maximum accuracy \green{drop} $Ad_{max}$ and a minimal latency $Lat_{std}$ imposed by the \gls{tml} environment.
The optimization is formalized in~\cref{eq:genetic_opt}, where we define our objective function as a joint minimization problem of the accuracy \green{drop} $F_{acc}$ of the decomposed \gls{cnn} and the latency of the accelerator $F_{lat}$.
To ensure the validity of each solution $x$, we also constrain our optimization based on a maximum accuracy \green{drop} of $Ad_{max}$ and a minimum latency $Lat_{std}$.
For a given solution $x$:

\begin{equation}
    \begin{split}
    &\textrm{minimize}\ \ \ (F_{acc}(x), F_{lat}(x)),\\
    &\phantom{\textrm{minimize:}}\ \ \ x = (p,z,e,m,s_w) \in (P, Z, E, M, S_{W}) \\ 
    &\textrm{s.t.\quad\quad} F_{acc}(x) - Ad_{max} \leq 0  \\
    &\phantom{\textrm{s.t.\quad\quad}} F_{lat}(x) - Lat_{std} \leq 0
    \end{split}
    \label{eq:genetic_opt}
\end{equation}

In order to tackle the optimization problem defined in \cref{eq:genetic_opt}, we employ a genetic algorithm, specifically the popular \gls{nsga}~\cite{deb_fast_2002}.
Evolutionary-based exploration enables the effective traversal of complex design spaces (as in our case, see \cref{subsec:dse}) in a reasonable amount of time, as they can be massively parallelized, thereby expediting the overall execution.
\gls{nsga} in particular, due to its elitist nature, ensures that the best performing (i.e., fittest) solutions are retained throughout the optimization process, and a front of Pareto-optimal accuracy-latency trade-offs can be obtained upon convergence.

% We base this choice on the fact that as aforementioned in~\cref{subsec:dse} the design space is extensive and difficult for approaches such as grid-search to navigate. 
% Additionally, \gls{nsga} uses an elitist selection process of the best $x$ guided by our constraints and results in a pareto front of optimal accuracy-latency trade-offs. 

Candidate solutions (i.e., chromosomes) represent discrete samples of our design space, and are encoded in integer values.
Starting from a randomly initialized population, the \gls{nsga} flow involves a sequence of fitness evaluations, uniform crossover, and random mutation operations, which guide the evolution to higher-quality solutions. 
% The first step of our optimization process is to sample candidate solutions $x$ from our design space.
Fitness evaluation follows the formulation in \eqref{eq:genetic_opt}.
\green{Note, that for our genetic search we use only a small subset of the test set and reserve the majority of the test set to evaluate the final accuracy of our framework solutions.} 
% Finally, we calculate our fitness functions used in our \gls{ga}. 
For ensuring exact accuracy calculations, we execute inference upon the approximate and deconstructed convolutional layers. 
First, we apply \gls{wmd} to the target \gls{cnn} layers individually, according to the decoded chromosome parameters (i.e., $P_h$ and $P_s$ of \cref{subsec:dse}).
% In order to holistically evaluate the impact of each parameters on the final \gls{cnn} accuracy, 
The resulting set of decomposed matrices $\mathcal{F}$ is then used to reconstruct the approximate convolutional layers and the \gls{po2}-based \gls{cnn}, following the reverse procedure, and allowing for directly executing inference with the approximate weights.
To determine $F_{lat}$, we consider our high-level latency surrogate model described in \cref{subsec:latency} to accurately quantify the total cycles using \cref{eq:latency_model}.
% and \eqref{eq:cost_f1}. \textcolor{red}{\textbf{Shouldn't this be ``using \cref{eq:latency_model}'' ?}}
Importantly, latency estimation is preceded by the iterative mapping algorithm used to arrange our \glspl{pe} into the systolic architecture.
This step utilizes our resource model (see \cref{subsec:resource_model}) to maximize utilization and deliver the \green{lowest} possible latency allowed by the selected set of \gls{wmd} parameters.

Overall, our genetic-based exploration enables resource-aware co-optimization of \gls{po2}-based \glspl{cnn} with programmable accelerators, generating Pareto-optimal trade-offs between bounded inference accuracy and latency.

% \input{tables/results.tex}

% follow the procedure described in~\cref{subsec:latency}, where the cost of our\gls{pe} can be calculated using~\cref{eq:cost_fgen} and \cref{eq:cost_f1}, then using the latency model~\cref{eq:latency_model} the final arrangement of the accelerator as well as the best latency can be determined using~\cref{alg:pe_mapping}. 

% \section{Experimental Results (FAU \& KIT) 1 page}
\fh[inline]{\small TO DO: Please search entire document, also table(s) footnotes, figures for:\\
-- SW\\[.25em]
-- $S_W$\\[.25em]
-- $S_w$\\[.25em]
-- $s_w$\\[.25em]
and unify! I guess at most two variants are required, or?}
\section{Results and Analysis}
\label{sec:experiments}
% Experiments and Results: Test bench for convolutional layers results from FPGA implementations\\
% Discussion: Comparison with standard MAC implementations, similar accelerator(s)
% \subsection{$\mathcal{F}$ Matrices Constraints Validation}

\green{This section shows the evaluation of our framework, reporting the solutions obtained for an exploration run with constraints taken from an 8-bit \gls{sa} using multipliers. We describe the benchmark and tools used for our experiments in \cref{subsec:experimentalsetup}, and the results obtained in \cref{subsec:evaluation}.
To evaluate our programmable accelerator architecture with WMD PEs considering hardware performance metrics, we also compare the obtained solutions against post-training, hardware-optimization techniques: \gls{ptq} with varying weight bit resolutions in \cref{subsec:ptq} and a state-of-the-art FPGA implementation of another Po2 approach in \cref{subsec:shiftcnn}}.

\subsection{Experimental Setup} \label{subsec:experimentalsetup}

We evaluate our framework on the set of \glspl{cnn} available in MLPerfTiny~\cite{banbury_mlperf_2021}, a widely used
\gls{tml} benchmark that \green{comprises representative CNN workloads customized to the stringent TinyML resource constraints.}
As \gls{cnn} benchmarks, it includes ResNet, MobileNetV1, and DS-CNN, trained on CIFAR-10, VWW, and Speech Commands, respectively. For \green{the framework evaluation}, we exclusively use the pre-trained models of all three \glspl{cnn} from MLPerfTiny, without additional \green{retraining or finetuning}.
\texttt{TFLite} is used to evaluate the accuracy of both our decomposed and the baseline \glspl{cnn}.
We evaluate \green{the solutions found with} our framework against
\green{traditional state-of-the-art \gls{sa}-based accelerators~\cite{jouppi_-datacenter_2017} that use low precision MAC units (i.e., weights from 4 to 8 bits)} across all MLPerfTiny \glspl{cnn}.
\blue{
The mapping of each \gls{cnn} to these \glspl{sa} is obtained using \cref{alg:pe_mapping}. Each row and column of the \gls{sa} can be supplied with weights and activations in a single clock cycle through appropriately instantiated \gls{bram}-based input and output buffers.
}
% \blue{
% This SA-based accelerator is obtained by instantiating a PE computing a single MAC operation instead of our custom PE in the same architecture from Figure \ref{fig:accelerator_scheduling}. The number of PEs instantiated is defined with Algorithm \ref{alg:pe_mapping} as well, so to utilize most of the available FPGA LUTs. As a design decision to allow enough BRAM bandwidth to feed the systolic array and decrease underutilization of the BRAMs capacity, weights are distributed in just enough BRAMs to feed one full row of PEs in one clock cycle. In a similar way, input activations are distributed such as one BRAM can feed the top PE of each column. The delays for weights updates and initial activations propagations are considered in the simulation of both \orange{baseline} and \frameworkdecomposition{} accelerators.
% }
% \blue{as existing po2-based approaches (e.g \cite{gudovsMobileNetkiy_shiftcnn_2017}) would violate our stringent resource constraints.}
% and the \blue{and the deployment accuracy is then validated by converting the decomposed \gls{cnn} to the \texttt{TFLite} format}.
% The \gls{wmd} parameters are then applied to each \gls{cnn} layer to obtain the $F$ matrices. These, are then re-combined and applied to the \texttt{Keras} model for the accuracy evaluation.
As our target platform, we use the 
% \blue{resource-constrained}\todo{Isn't any FPGA resource-constrained? The FPGA is rather mid-size, while its package is very small with low pin count} 
\texttt{Artix-7 XC7A100TCSG324-1} \gls{fpga}. A fully parametrized \gls{sa} using our \gls{wmd} \glspl{pe} \green{as well as the \glspl{sa} using MAC units as PEs} are implemented in \texttt{SystemVerilog}.\fh{While only our \gls{wmd} \glspl{pe}? Aren't the same tools used to obtain results for both the baseline SAs and CoMD?}
\texttt{Vivado 2023.1} is used for simulation and synthesis to obtain \gls{fpga} resource utilization in terms of \glspl{lut}, \glspl{ff}, and \glspl{bram}, as well as to estimate power. For our resource model, we synthesize its basic blocks and extract their resource utilization in terms of \glspl{lut} (see \cref{subsec:resource_model}).
The final latency is calculated for each \gls{cnn} on a layer-wise basis using \texttt{Vivado} simulation, \green{from which toggle rates are also obtained to serve as inputs for the power consumption analysis.}
Functional verification uses uniformly random inputs, decomposed weights, and $8$-bit activations.

% For our final latency evaluation, we also need to account for the additional cycles introduced when filling the \gls{pe} weight registers.
% As mentioned in~\cref{subsec:systolic}, depending on the \gls{wmd} parameters we can allocate the required number of \glspl{bram} to pre-load the weights of one \gls{pe} row within a single cycle.
% This corresponds to an added latency, proportional to the number of rows every $O_{x,y}$ cycles.
% %In an effort
% To maintain a fair comparison, the same number of \glspl{bram} is allocated per  \gls{pe} of the \orange{baseline} \gls{sa}. 
% The latency can then be calculated individually for all layers of the target \gls{cnn}.
% Although the data traffic when processing the input activations is not memory-bound in the FPGA used for validation, feeding the weights is. 
Our \gls{nsga} is implemented using \texttt{pymoo}~\cite{blank_pymoo_2020}, with  an initial population of $250$ chromosomes, running for up to $20$ generations.
We set the crossover probability and \green{degree of mutation }$\eta_c$ to $0.9$ and $15$, respectively, and the mutation $\eta_m$ to $5$. \green{For the genetic exploration we reserve $10\%$ of the test set for the exploration itself and $90\%$ of the test set to evaluate the final accuracy of our designs.}
% \fh{Please check: $\eta$ is set once to 15 and once as \enquote{mutation $\eta$} to 5. Do they the two refer to the same?}
% As described in~\cref{subsec:dse}, we partition our design space into $P_s$ and $P_h$, exploring a range of values for each \gls{wmd} parameter. 
For our largest benchmark (i.e., MobileNet), the size of the explored design space---as defined by the \gls{wmd} parameters $P_h, P_s$---reaches \green{for $P_s=54$, $2^{12}\cdot 54=221\,184$} possible solutions, highlighting the need for employing genetic-based exploration within our framework.
% Specifically, for ResNet and DS-CNN we evaluate $6$ and $4$ layers, respectively, with $|P_s|,|P_h|=108,2$. This results in a search space of $4^2\cdot108=1\,728$ for DS-CNN and $6^2\cdot108=3\,888$ for ResNet.
% However, for MobileNet, we evaluate a total of $12$ layers.
% For a faster convergence, we set $P_s=54$ for a total of $2^{12}\cdot 54=221\,184$ possible solutions.
% This highlights the need for using a \gls{ga} to traverse the design space of deeper \glspl{cnn} effectively.
\subsection{Evaluation} \label{subsec:evaluation}

\captionsetup[sub]{labelfont=normalfont,textfont=normalfont}

% ---------- Figure 1: Pareto (half-column) ----------
\begin{figure}[b]
    \centering
    \includegraphics[width=\linewidth]{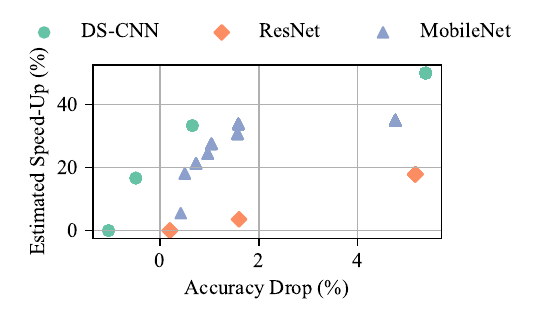}\\[-1em]
    \caption{Pareto front obtained from NSGA-II exploration for different network models showing \green{accuracy loss and normalized speed-ups $S = Lat_{\mathrm std} /Lat(x)$ for each found Pareto point $x$.}}
    \label{fig:pareto_halfcol}
\end{figure}

% ---------- Figure 2: PTQ (half-column), latency top + accuracy bottom ----------
\begin{figure}[t]
    \centering
    (a) PTQ results for DS-CNN\\[.5em]
    \includegraphics[trim=1mm 0mm 160mm 7mm, clip, height=31mm]{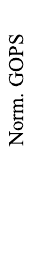}\includegraphics[trim=6mm 0mm 110mm 6.5mm, clip, height=31mm]{figures/ptq_latency_3models.pdf}\hspace{8mm}\includegraphics[trim=1mm 0mm 156mm 4mm, clip, height=31mm]{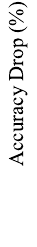}\includegraphics[trim=7mm 0mm 106mm 2.5mm, clip, height=31mm]{figures/ptq_accuracy_3models.pdf}\\[.25em]
    (b) PTQ results for ResNet\\[.5em]
    \includegraphics[trim=1mm 0mm 160mm 7mm, clip, height=31mm]{figures/ptq_latency_3models.pdf}\includegraphics[trim=60.5mm 0mm 55.5mm 6.5mm, clip, height=31mm]{figures/ptq_latency_3models.pdf}\hspace{8mm}\includegraphics[trim=1mm 0mm 156mm 4mm, clip, height=31mm]{figures/ptq_accuracy_3models.pdf}\includegraphics[trim=61.5mm 0mm 51.5mm 2.5mm, clip, height=31mm]{figures/ptq_accuracy_3models.pdf}\\[.25em]
    (c) PTQ results for MobileNet\\[.5em]
    \includegraphics[trim=1mm 0mm 160mm 7mm, clip, height=31mm]{figures/ptq_latency_3models.pdf}\includegraphics[trim=115.mm 0mm 1mm 6.5mm, clip, height=31mm]{figures/ptq_latency_3models.pdf}\hspace{8mm}\includegraphics[trim=1mm 0mm 156mm 4mm, clip, height=31mm]{figures/ptq_accuracy_3models.pdf}\includegraphics[trim=116mm 0mm -3mm 2.5mm, clip, height=31mm]{figures/ptq_accuracy_3models.pdf}\\[-1mm]
    \caption{Post-training quantization (PTQ) comparison for (a) DS-CNN, (b)~ResNet, and (c) MobileNet. The graphs show the normalized GOPS on the left and the accuracy drop on the right.}
    \label{fig:ptq_halfcol}
\end{figure}

\begin{table*}[ht]
\setlength\tabcolsep{5.1pt}
\centering
\caption{Detailed comparison of DS-CNN from our proposed framework against systolic arrays of different weight resolution.}
\begin{threeparttable}
\begin{tabular}{l|ccccc|ccc}
\toprule
\textbf{} &
  \multicolumn{5}{c|}{\textbf{MAC-based \gls{sa}}} &
  \multicolumn{1}{c}{\textbf{Ours}} \\
\midrule
% \midrule
\textbf{Weight Resolution} &
  \textbf{4-bit} &
  \textbf{5-bit} &
  \textbf{6-bit} &
  \textbf{7-bit} &
  \textbf{8-bit} &
  \textbf{8-bit\tnote{1}} \\

\textbf{Top-1 Accuracy (\%)} &
80.83 & 91.05 & 91.67 & \textbf{92.36} & 91.95 & 
{90.63}
\\
\textbf{Number of LUTs \green{(overall utilization)}} &
62531 (99\%) & 62586 (99\%) & 62825 (99\%) & 61873 (99\%) & 61612 (97\%) &
{59922 (95\%)} 
\\
\textbf{Number of 36-Kb BRAMs \green{(overall utilization)}} &
18 (13\%) & 18.5 (14\%) & 37.5 (28\%) & 13.5 (10\%) & {7.5 (6\%)} &
{11 (8\%)}
\\
\textbf{Number of FFs \green{(overall utilization)}} &
32301 (25\%) & 27578 (22\%) & 25900 (20\%) & 21233 (17\%) & {22927 (18\%)} &
{19033 (15\%)}
\\
\midrule
\textbf{Frequency (MHz)} &
\textbf{125} & 113 & 122 & 111 & 114 & 
{122} \\
\textbf{Latency ($\bm{\mu}$s)} &
21.02 & 22.99 & 31.34 & 32.93 & 30.79 & 
\textbf{16.88} \\
\textbf{Peak Throughput (GOPS)} &
156 & 116 & 109 & 81 & 89 & 
\textbf{187}
\\
\textbf{Estimated Power (mW)} &
1166 & 1117 & 1067 & \textbf{1060} & 1097 &
{1237}
\\
\textbf{Estimated Energy ($\bm{\mu}$J)} &
24.51 & 25.68 & 33.43 & 34.91 & 33.77 &
\textbf{20.88}
\\
\textbf{Estimated \green{Energy Efficiency (GOPS/W)}} &
133.59 & 104.00 & 102.37 & 76.36 & 81.13 &
\textbf{151.17}
\\
\bottomrule
\end{tabular}

\begin{tablenotes}\scriptsize
\item[]
\textsuperscript{1}\,PW-Conv(1–4), $P$=2, $Z$=3, $E$=3, $M$=4, $S_W$=4
\end{tablenotes}

\end{threeparttable}
\label{tab:results_ptq_dscnn}
\end{table*}

\begin{table*}[ht]
\setlength\tabcolsep{5.1pt}
\centering
\caption{Detailed comparison of ResNet from our proposed framework against systolic arrays of different weight resolution.}
\begin{threeparttable}
\begin{tabular}{l|ccccc|ccc}
\toprule
\textbf{} &
  \multicolumn{5}{c|}{\textbf{MAC-based \gls{sa}}} &
  \multicolumn{1}{c}{\textbf{Ours}} \\
\midrule

\textbf{Weight Resolution} &
  \textbf{4-bit} &
  \textbf{5-bit} &
  \textbf{6-bit} &
  \textbf{7-bit} &
  \textbf{8-bit} &
  \textbf{8-bit\tnote{2}} \\

\textbf{Top-1 Accuracy (\%)} &
53.55 & 75.31 & 85.16 & 86.47 & \textbf{87.08} & 
85.63
\\
\textbf{Number of LUTs \green{(overall utilization)}} &
62531 (99\%) & 62586 (99\%) & 62415 (98\%) & 61873 (98\%) & 60757 (96\%) &
{55450 (87\%)} 
\\
\textbf{Number of 36-Kb BRAMs \green{(overall utilization)}} &
18 (13\%) & 18.5 (14\%) & 13 (10\%) & 13.5 (10\%) & {13.5 (10\%)} &
{54 (40\%)}
\\
\textbf{Number of FFs \green{(overall utilization)}} &
32301 (25\%) & 27578 (22\%) & 25427 (20\%) & 21233 (17\%) & {22618 (18\%)} &
{13674 (11\%)}
\\
\midrule
\textbf{Frequency (MHz)} &
125 & 113 & \textbf{123} & 111 & 113 & 
{114} \\
\textbf{Latency ($\bm{\mu}$s)} &
\textbf{236.80} & 259.74 & 278.12 & 306.82 & 302.58 & 
{250.24} \\
\textbf{Peak Throughput (GOPS)} &
\textbf{155.77} & 116.17 & 111.39 & 80.95 & 84.57 & 
{117}
\\
\textbf{Estimated Power (mW)} &
1166 & 1117 & 1064 & 1060 & \textbf{1041} &
{1709}
\\
\textbf{Estimated Energy ($\bm{\mu}$J)} &
\textbf{276.11} & 290.13 & 295.92 & 325.23 & 314.98 &
427.66
\\
\textbf{Estimated \green{Energy Efficiency (GOPS/W)}} &
\textbf{133.59} & 104.00 & 104.69 & 76.36 & 81.24 &
68.46
\\
\bottomrule
\end{tabular}

\begin{tablenotes}\scriptsize
\item[]
\textsuperscript{2}\,Conv3x3(1–7), $P$=2, $Z$=3, $E$=3, $M$=16, $S_W$=4
\end{tablenotes}

\end{threeparttable}
\label{tab:results_ptq_resnet}
\end{table*}

\begin{table*}[ht]
\setlength\tabcolsep{5.1pt}
\centering
\caption{Detailed comparison of MobileNet from our proposed framework against systolic arrays of different weight resolution.}
\begin{threeparttable}
\begin{tabular}{l|ccccc|ccc}
\toprule
\textbf{} &
  \multicolumn{5}{c|}{\textbf{MAC-based \gls{sa}}} &
  \multicolumn{1}{c}{\textbf{Ours}} \\
\midrule

\textbf{Weight Resolution} &
  \textbf{4-bit} &
  \textbf{5-bit} &
  \textbf{6-bit} &
  \textbf{7-bit} &
  \textbf{8-bit} &
  \textbf{8-bit\tnote{3}} \\

\textbf{Top-1 Accuracy (\%)} &
78.33 & 81.51 & 83.11 & \textbf{84.18} & {84.07} & 
{82.88}
\\
\textbf{Number of LUTs \green{(overall utilization)}} &
62531 (99\%) & 62586 (99\%) & 62959 (99\%) & 61873 (98\%) & 62367 (98\%) &
{62506 (99\%)} 
\\
\textbf{Number of 36-Kb BRAMs \green{(overall utilization)}} &
18 (13\%) & 18.5 (14\%) & 15.5 (11\%) & 13.5 (10\%) & {15 (11\%)} &
{37.5 (27.7\%)}
\\
\textbf{Number of FFs \green{(overall utilization)}} &
32301 (25\%) & 27578 (22\%) & 25545 (20\%) & 21233 (17\%) & {23170 (18\%)} &
{18390 (17.5\%)}
\\
\midrule
\textbf{Frequency (MHz)} &
\textbf{125} & 113 & 123 & 111 & 113 & 
{114} \\
\textbf{Latency ($\bm{\mu}$s)} &
{100.34} & 115.05 & 124.59 & 149.65 & 147.99 & 
\textbf{87.20} \\
\textbf{Peak Throughput (GOPS)} &
{155.77} & 116.17 & 112.37 & 80.95 & 86.83 & 
\textbf{161}
\\
\textbf{Estimated Power (mW)} &
1166 & 1117 & 1086 & \textbf{1060} & {1073} &
{1717}
\\
\textbf{Estimated Energy ($\bm{\mu}$J)} &
\textbf{116.99} & 128.52 & 135.30 & 158.63 & {158.79} &
{149.72}
\\
\textbf{Estimated \green{Energy Efficiency (GOPS/W)}} &
\textbf{133.59} & 104.00 & 103.47 & 76.36 & 80.92 &
{93.77}
\\
\bottomrule

\end{tabular}

\begin{tablenotes}\scriptsize
\item[]
\textsuperscript{3}\,PW-Conv(2–13), $P$=2, $Z$=3, $E$=3, $M$=8, $S_W$=4
\end{tablenotes}

\end{threeparttable}
\label{tab:results_ptq_mobilenet}
\end{table*}

First, we evaluate the effectiveness of our proposed framework in delivering Pareto-optimal solutions in terms of \green{classification accuracy drop $F_{\mathrm acc}(x)$ in percent using the accuracy of the corresponding floating-point trained network model, and normalized speedups $S = Lat_{\mathrm std} /Lat(x)$ for each found Pareto point $x$ with $Lat_{\mathrm std}$ corresponding to the latency of an 8-bit baseline SA.}
%classification accuracy loss (\green{$F_{\mathrm acc}$ with respect to the floating-point trained neural network model}) and \green{latency $F_{\mathrm lat}$(with respect to the $n$-bit baseline \gls{sa} and the WMD PEs accelerator.)}
Figure~\ref{fig:pareto_halfcol} presents the obtained Pareto fronts from our \gls{nsga}-based exploration across all studied benchmarks.

Overall, we observe a wide range of trade-offs between latency and accuracy, covering both ends of the spectrum---from solutions that prioritize high accuracy to those achieving significant latency gains.
% The results of our co-exploration of the three evaluated \glspl{cnn} can be seen in~\cref{fig:nsga_pareto}, demonstrating the efficacy of our \frameworkdecomposition{} in finding a wide range of solutions for accelerating \gls{cnn} inference. 
% Notably, the Pareto front provides solutions with varying accuracy and speed-up trade-offs.
Most of the \green{found} Pareto-optimal solutions fall within the region of a conservative 2\% accuracy drop, reaching up to \red{34\%} \blue{estimated} speed-up over the \orange{baseline}.
Interestingly, our framework discovered a decomposed \gls{po2}-based version of DS-CNN, which improves its accuracy by 1\% during an iso-latency evaluation.
This demonstrates the effectiveness of our framework in exploring potential decomposed \glspl{cnn} and accelerator pairs, enabling \gls{po2} inference with small accuracy loss and notable latency gains.

% \input{figures/nsga_pareto.tex}
% \subsection{Synthesis Results}

Next, we perform an in-depth analysis of selected solutions from each Pareto front of \cref{fig:pareto_halfcol} in terms of accuracy, hardware utilization, \blue{latency}, throughput, \blue{and power}, for the targeted \gls{fpga} board.
Specifically, we select the highest-performing decomposed \gls{cnn} with an accuracy loss below the 2\% threshold. 
\green{\cref{tab:results_ptq_dscnn}, \cref{tab:results_ptq_resnet}, and \cref{tab:results_ptq_mobilenet}} present the obtained results in comparison with the \green{$n$-bit \glspl{sa}} across all benchmarks. \green{While this section focuses on the 8-bit \gls{sa} comparison against the WMD PEs accelerator solutions obtained with its latency constraint as input to our framework, a detailed comparison with all bit-width \glspl{sa} variants is discussed in \cref{subsec:ptq}.}

Our selected solutions feature a considerable speed-up of \red{1.55$\times$} on average, with an average accuracy loss of \red{1.3\%}.
Notably, our \gls{po2}-based solution for DS-CNN achieves \red{1.82$\times$} speed-up with only \red{1.15\%} accuracy drop, showcasing a favourable trade-off.
% With respect to resource utilization, our \frameworkdecomposition{} is able to obtain performance gains with much fewer \glspl{lut}---\red{xxx\%} on average---compared to the \gls{sa} \orange{baseline}, which nearly utilizes the entirety of available resources. 
% \blue{
% We also observe increased \gls{bram} and \gls{ff} utilization, which is tied to increased bandwidth requirements of our increased parallelization. Note, that according to~\cref{tab:results} the main limiting factor of the \gls{sa} implementation is the number of \glspl{lut}. By applying our \frameworkdecomposition{}, we lift this constraint, while taking advantage of available \glspl{bram} and \glspl{ff} that would otherwise remain idle. The increase in \glspl{ff} results from additional pipeline stages within the \gls{wmd} \gls{pe} and the higher degree of parallelization enabled by our approach. Likewise, the elevated \gls{bram} utilization is due to the increased parallelization attainable by our approach. The positive impact of our resource-efficiency is also reflected by an average throughput gain of \red{27\%}
% }
\blue{
Our lightweight \gls{pe} and \gls{sa} architecture enable higher degrees of parallelism (i.e., more instantiated \glspl{pe} within the same fabric and using a similar number of \glspl{lut}), leading to increased throughput.
The positive impact of the achieved resource efficiency is reflected by an average throughput gain of \red{41\%}.
The increased throughput is accompanied by higher bandwidth demands, reflected in increased \gls{bram} utilization and additional \glspl{ff} due to the deeply pipelined microarchitecture of our \gls{pe}. However, the main limiting factor of the \orange{baseline} \glspl{sa} implementation remains the number of \glspl{lut} and not \glspl{bram} and \glspl{ff}.
Overall, our approach achieves performance improvements under the stringent energy and resource constraints of the \gls{tml} domain. 
}
% \input{tables/wmd_cnns}
% The chosen \gls{wmd} parameters for each solution of \cref{tab:results}.
% As noted in~\cite{lehnert2023most}, higher $M/S_W$ ratios generally improve accuracy.
% DS-CNN stands out with the lowest ratio of $1$, highlighting its robustness to the \gls{wmd} process.

\begin{figure}[t]%
%     \centering
\hspace{-4mm}\includegraphics[width=1.06\linewidth]{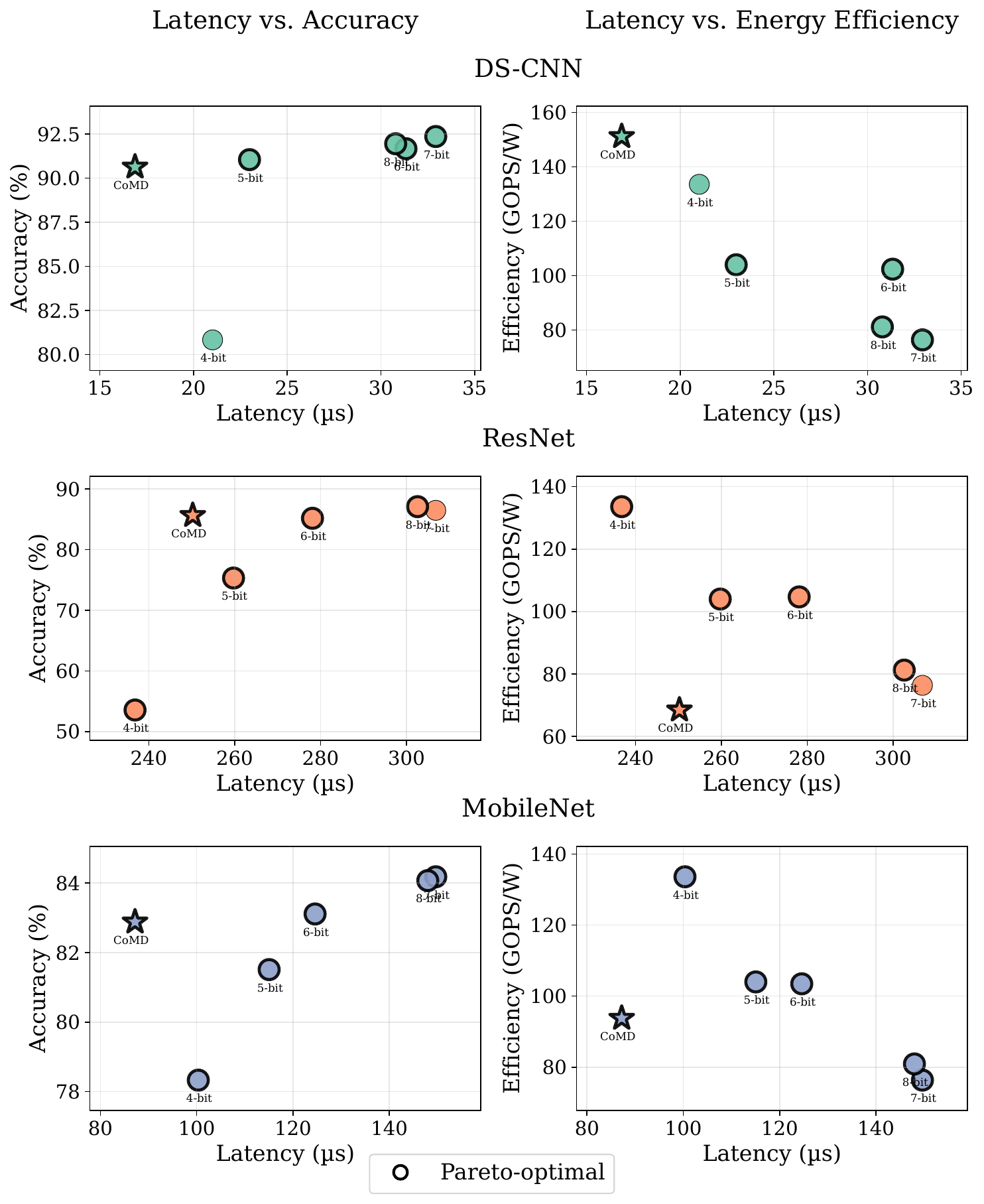}
    \caption{Comparison of \gls{sa} and obtained accelerators' performance metrics showing latency, accuracy, and energy efficiency trade-offs across models.}
    \label{fig:ptq_pareto}
\end{figure}

\green{ \subsection{Comparison Against State-of-the-Art Low-Precision MAC-based Systolic Array}} \label{subsec:ptq}

We compare our approach against \gls{ptq}, a state-of-the-practice technique for improving hardware efficiency~\cite{gholami2022survey}. Using the same \orange{baseline} \gls{sa} architecture and adjusting the precision of the \gls{mac} unit, we progressively quantize the weights of the evaluated \glspl{cnn} to lower bit-widths while keeping activations fixed at $8$~bits. \Cref{fig:ptq_halfcol} shows the achievable peak throughput of the \gls{mac}-based \gls{sa} at various bit-widths, normalized with respect to the accelerators obtained with our framework. 
Overall, we observe that \gls{ptq} at bit-widths lower than $5$~bits incurs substantial accuracy loss, well over the $2\%$ accuracy \green{drop} threshold.
At the same time, our approach consistently outperforms \gls{ptq} within the $2\%$ accuracy \green{drop} threshold
showcasing the effectiveness of our \gls{po2}-based framework in providing throughput improvements at a minimal loss in terms of accuracy. \green{\green{\cref{tab:results_ptq_dscnn}, \cref{tab:results_ptq_resnet}, and \cref{tab:results_ptq_mobilenet}} contain the PTQ obtained top-1 accuracies, hardware resources and performance metrics for the different weight resolution systolic arrays. The best values for each row are highlighted for an easier comparison.  The 4-bit SA, although providing advantages in many hardware performance metrics, causes a top-1 accuracy drop of at least 6\% for the analyzed CNNs, which is why this configuration is excluded in the following comparisons.
It can be observed that the DS-CNN obtained configuration outperforms the \glspl{sa} across most of the hardware performance metrics.
In general, the slightly higher estimated power consumption of the obtained accelerators is compensated by the higher peak throughput\fh[inline]{Why \enquote{peak throughput} and not just \enquote{throughput}? The reader may ask what the real throughput reached in the experiments is. On a GPU or CPU peak performance always has the negative touch that it is very difficult to reach in an implementation. Can we omit the \enquote{peak} throughout the paper?}, which leads to lower latencies and higher energy efficiency in terms of GOPS/W. Figure~\ref{fig:ptq_pareto} shows that the accelerator solutions found with our framework lie in the Pareto front of all evaluated CNNs, demonstrating optimal trade-offs between latency, accuracy, and energy efficiency.}

\green{Considering the \glspl{sa} resolutions with closer top-1 accuracy to the analyzed  configurations obtained with our framework (5-bit for DS-CNN, 6-bit for ResNet, and 6-bit for MobileNet), an average speed-up of \red{1.3$\times$} is obtained, highlighting the main advantage of our approach. Additionally, power and energy savings of 10.74\% and 18.67\%, respectively, are obtained for DS-CNN.}

\green{ \subsection{Comparison Against State-of-the-Art Post-Training Po2 FPGA Implementation} \label{subsec:shiftcnn}}
%\nopagebreak[4]
Finally, we compare our framework methodology to the state-of-the-art \gls{po2} acceleration technique, ShiftCNN, proposed by \citeauthor{gudovskiy_shiftcnn_2017}~\cite{gudovskiy_shiftcnn_2017}.
\green{We re-implement ShiftCNN targeting the same FPGA as in our work and by precisely following the design directives detailed in~\cite{gudovskiy_shiftcnn_2017}}. \green{This approach approximates a normalized weight by adding $N$ numbers from a set of negative powers of 2, each selected by a $B$-bit index.
Then, the hardware architecture includes a precomputed tensor storing right-shifted input activations by different amounts, for which just one is selected using encoded $N \cdot B-$bit weights as inputs to multiplexers. A parameter $C$ is also considered in the architecture to define the number of weight and activation pairs that the accelerator can receive as inputs per cycle, which at the same time defines the number of multiplexers instantiated. While not explicitly described, $N \cdot C$ multiplexers are needed to keep the multiplication result in a single cycle. All multiplexers' outputs are accumulated with an adder tree.}

% ---------- Figure 3: Shift-CNN (half-column), latency left + accuracy right ----------
\begin{figure}[b]
    \centering
    \includegraphics[width=\linewidth]{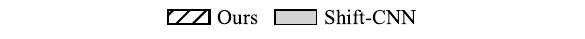}\\[-0mm]
    \begin{minipage}[t]{0.49\linewidth}
        \hspace{-2mm}\includegraphics[width=1.03\linewidth]{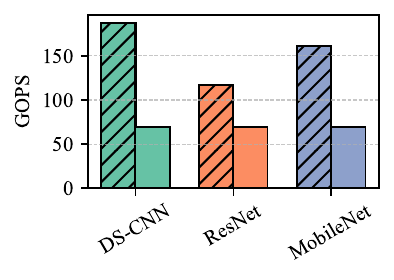}
    \end{minipage}
    \hfill
    \begin{minipage}[t]{0.49\linewidth}
        \centering
        \includegraphics[width=1.03\linewidth]
        {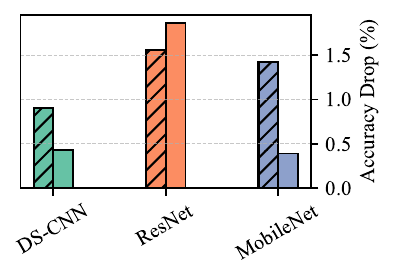}
    \end{minipage}
    \caption{Comparison between our framework and Shift-CNN: GOPS (left) and accuracy drop (right). \TODO{GM: Here we are not matching the acc, it could be said that there is another Pareto point for ShiftCNN that gets acc==same and better GOPS}}
    \label{fig:shiftcnn_halfcol}
\end{figure}

\begin{table*}[ht]
\setlength\tabcolsep{5.1pt}
\centering
\caption{Throughput and accuracy drop obtained by different ShiftCNN variants for the MLPerfTiny selected CNNs}
\begin{threeparttable}
\begin{tabular}{ccccccc|cccc}
\toprule
  \multicolumn{7}{c|}{\textbf{}} &
  \multicolumn{3}{c}{\textbf{Accuracy Drop}} \\
  \midrule

  \textbf{$N$} &
  \textbf{$B$} &
  \textbf{LUTs} &
  \textbf{Frequency (MHz)} &
  \textbf{Instantiable Adder Trees} &
  \textbf{OPs} &
  \textbf{GOPS} &
  \textbf{DS-CNN} &
  \textbf{ResNet} &
  \textbf{MobileNet} \\
\midrule
{4} & {2} & 11791 & 101 & 5 & 640 & {64.49} & {0.43} & {0.39} & {1.86}
\\
{3} & {3} & 13793 & 93  & 4 & 512 & {47.58} & {1.53} & {0.14} & {6.22}
\\
{3} & {2} & 9516 & 108 & 6 & 768 & {82.57} & {7.71} & {2.74} & {30.8}
\\
\bottomrule
\end{tabular}

% \begin{tablenotes}\scriptsize
% \item[]
% $^1$ PW-Conv(1–4), P=2, Z=3, E=3, M=4, SW=4
% $^2$ Conv3x3(1–7), P=2, Z=3, E=3, M=16, SW=4,
% $^3$ PW-Conv(2–13), P=2, Z=3, E=3, M=8, SW=4,
% \end{tablenotes}

\end{threeparttable}
\label{tab:results_shiftcnn}
\end{table*}

As can be seen in \cref{fig:shiftcnn_halfcol}, ShiftCNN incurs an average accuracy drop of \red{$0.9\%$} across all evaluated \glspl{cnn}. \green{When setting the parameters $N=2$, $B=4$ and $C=128$ with 8-bit input activations as used in~\cite{gudovskiy_shiftcnn_2017} results for the ShiftCNN accelerator,} our co-design methodology outperforms \green{it} by an average of \red{2.4$\times$} in terms of throughput, while still remaining below the $2\%$ accuracy \green{drop} threshold. \green{Our accelerators advantage comes from the higher number of GOPS obtained with similar achievable frequency and LUT resources. For instance, using $N=4$ and $B=2$ to keep the accuracy drop constraint of 2\% for all evaluated CNNs, ShiftCNN reaches 64.49 GOPS while the lowest value for our obtained accelerators is 87.20 GOPS for MobileNet. \cref{tab:results_shiftcnn} shows different parameter values for the ShiftCNN architecture with the corresponding throughput and accuracy obtained for the MLPerfTiny CNNs selected.
The LUTs column reports the synthesis results for a single adder tree, from which the total number of instances is obtained, utilizing the same available FPGA LUTs as for the accelerators obtained with our framework.
By comparing with the obtained configurations detailed in \cref{tab:results_ptq_dscnn}, \cref{tab:results_ptq_resnet}, and \cref{tab:results_ptq_mobilenet}, it can be seen that the increased throughput that may be obtained for ShiftCNN parameters that reduce required LUTs cannot reach the values of our accelerators. For instance, using the parameters $N = 3$, $B=2$ helps ShiftCNN to reach a throughput of 82.57~GOPS, which at the cost of high accuracy drops, is still way below the obtained throughputs of our accelerators. } \\

\section{Conclusions}

% make more positive put results in
Despite the substantial benefits of multiplier-less \gls{po2} acceleration for the \gls{tml} domain, its implementation remains challenging. Training-based approaches offer high accuracy but are impractical when the training dataset is unavailable. On the other hand, existing post-training approaches for achieving \gls{po2} inference do not take into account the underlying hardware architecture, resulting in a high resource utilization.

%\green{Considering all these issues,} we propose a novel post-training co-design methodology for accelerating tiny \glspl{cnn} using \gls{po2} via approximate \gls{wmd}, providing accuracy-latency trade-offs and enabling low-latency \gls{cnn} inference with a minimal drop in accuracy.}
\textcolor{fhcolor}{To address these limitations, we proposed a novel post-training co-design methodology that accelerates tiny \glspl{cnn} using \gls{po2} representations via approximate \gls{wmd}.
Our approach explicitly incorporates hardware awareness, enabling flexible accuracy–latency trade-offs and achieving low-latency \gls{cnn} inference with only minimal accuracy degradation.} \green{From evaluations using the PerfTinyML benchmark CNNs, we obtained an average of \red{33\%} latency improvement at an average accuracy loss of \red{1.3\%} compared to typical systolic-array-based FPGA accelerators.
When compared with post-training-quantization hardware-optimized variants, the solutions obtained by our framework lie on the Pareto front of the accuracy drop and latency objectives used for the exploration.
Additionally, a comparison with a state-of-the-art that also employs Po2 decomposition for CNN acceleration shows an average $2.4\times$ increase in achievable throughput with our hardware solutions, considering an accuracy drop of less than 2\% across the analyzed CNNs.}
% We design a configurable \gls{po2} accelerator, our \frameworkdecomposition{} then collaboratively searches the accelerator-\gls{wmd} design space using a \gls{ga} and guided by resource and latency models.
% Our \frameworkdecomposition{} framework enables accuracy-latency trade-offs, allowing for low-latency \gls{cnn} inference with a slight drop in accuracy.

% \section*{Acknowledgment} This work was partly supported by the Deutsche Forschungsgemeinschaft (DFG, German Research Foundation) under project number 524986327 (NA$^3$Os).

%
% \bibliographystyle{IEEEtran}
% \bibliography{references}
\printbibliography[title=References]

% Biography text here without a photo.

% \begin{IEEEbiography}[{\includegraphics[width=1in,height=1.25in,clip,keepaspectratio]{fig1.png}}]{IEEE Publications Technology Team}
% In this paragraph you can place your educational, professional background and research and other interests.\end{IEEEbiography}

\end{document}